\documentclass[preprint,aps,prb,showpacs,amsmath]{revtex4}
\usepackage{latexsym}
\usepackage{amssymb}
\usepackage{graphicx}
\usepackage{bm}
\usepackage[english]{babel}
\usepackage[latin1]{inputenc}
\usepackage{amsmath}
\usepackage{amsfonts}

\newcommand{\bq}{\begin{equation}}
\newcommand{\eq}{\end{equation}}
\newcommand{\bn}{\begin{eqnarray}}
\newcommand{\en}{\end{eqnarray}}

\begin{document}

\title{Positive current noise cross-correlations in capacitively coupled double quantum dots with ferromagnetic leads}

\author{Bing Dong}
\affiliation{Department of Physics, Shanghai Jiaotong University, 800 Dongchuan
Road, Shanghai 200240, China}

\author{X. L. Lei}
\affiliation{Department of Physics, Shanghai Jiaotong University, 800 Dongchuan
Road, Shanghai 200240, China}

\author{N. J. M. Horing}
\affiliation{Department of Physics and Engineering Physics, Stevens Institute of Technology, Hoboken, New Jersey 07030, USA}

\begin{abstract}

We examine cross-correlations (CCs) in the tunneling currents through two parallel interacting quantum dots coupled to four independent ferromagnetic 
electrodes. We find that when either one of the two circuits is in the parallel configuration with sufficiently strong polarization strength, a new 
mechanism of dynamical spin blockade, i.e., a spin-dependent bunching of tunneling events, governs transport through the system together with the 
inter-dot Coulomb interaction, leading to a sign-reversal of the zero-frequency current CC in the dynamical channel blockade regime, and to 
enhancement of positive current CC in the dynamical channel anti-blockade regimes, in contrast to the corresponding results for the case of 
paramagnetic leads.

\end{abstract}


\maketitle

Quantum noise cross-correlation (CC) in multiterminal mesoscopic devices far from equilibrium has recently become an active issue because it 
characterizes the degree of correlation and the statistics of the charge carriers.\cite{Blanter,Nazarov} It is believed that in a noninteracting 
system, the current CC between different normal-metallic leads is always negative due to the fermionic statistics of electrons;\cite{Buttiker} this 
has been confirmed experimentally in a Hanbury Brown-Twiss setup.\cite{HBT} On the other hand, much theoretical work has predicted that current CC 
may become positive in the following situations: for a hybrid superconductor-normal system;\cite{Anantram} for a system with three-terminal 
ferromagnetic leads;\cite{Cottet} also for a system in the Coulomb-interaction mediated regime;\cite{Martin} and as a result of feedback effects of 
external voltage fluctuations.\cite{Wu}  

Very recently, Coulomb-interaction-induced positive current CC was experimentally observed in a double quantum dot (QD) system, with two parallel QDs 
coupled via an inter-dot Coulomb interaction, and also coupled to four independent electrodes and two independent gates, as shown in 
Fig.~1.\cite{McClure} It was found that when the system is biased on the specific occupation parameters, (1,0) and (0,1) [$(M, N)$ denotes the 
electron numbers in the top and bottom QDs], by means of tuning two gate voltages, the electrons occupying one dot during sequential tunneling 
through that QD can enhance either the tunneling-in rate or the tunneling-out rate of the other dot due to the Coulomb interaction (dynamical Coulomb 
anti-blockade), resulting in the appearance of a positive CC between the currents through the two QDs (i.e. electron bunching).\cite{McClure,Haupt} 
However, the current CC still remains negative in the transport regimes, (0,0) and (1,1), due to the dynamical Coulomb blockade effect (electron 
anti-bunching). 

In this Letter, we analyze the current CC for the same structure as the one studied in Refs.~\onlinecite{McClure,Haupt}, but with ferromagnetic 
leads. Experiments with a strongly interacting QD and ferromagnetic contacts have recently been performed successfully,\cite{fif} and it is not 
difficult to make such a structure with parallel-coupled QDs connected to four independent ferromagnetic leads, as shown in Fig.~1.
Our main finding is that sufficiently polarized contacts can lead to positive current CC even in the Coulomb blockaded region (0,0) if only one of 
the two circuits is in the parallel configuration. Similar to results for the three-terminal QD with ferromagnetic electrodes studied by Cottet {\it 
et al},\cite{Cottet} the present result also stems from dynamical spin blockade, associated with dynamical channel blockade. Furthermore, we find 
that the sign of current CC reduces to the result for paramagnetic electrodes with increasing spin relaxation rate.   

\begin{figure}[htb]
\includegraphics[height=5cm,width=4cm]{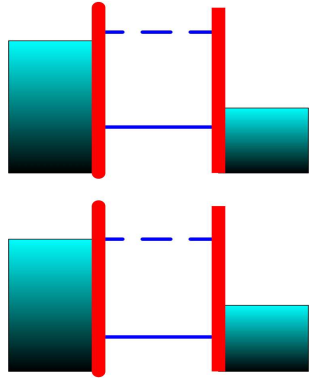}
\caption{(Color online) Schematic diagram of the system.}
\label{fig1}
\end{figure}

The system we study here consists of two parallel single-level QDs [top and bottom ($i=t,b$)] with energies $\varepsilon_i$ and inter-dot Coulomb 
interaction $U$, which are coupled to four collinearly spin-polarized leads with net spin-independent tunneling rates $\gamma_i$ and polarization 
strengths $p_i$ ($0\leq p_i <1$) (Fig.~1). Accordingly, the ferromagnetism of the leads in the top/bottom circuit can be accounted for by the 
spin-dependent tunneling rates: $\Gamma_{Li\uparrow}=\Gamma_{Ri\uparrow}=\gamma_i (1+p_i)$ and $\Gamma_{Li\downarrow}=\Gamma_{Ri\downarrow}=\gamma_i 
(1-p_i)$ for the parallel (P) configuration; $\Gamma_{Li\uparrow}=\Gamma_{Ri\downarrow}=\gamma_i (1+p_i)$ and 
$\Gamma_{Li\downarrow}=\Gamma_{Ri\uparrow}=\gamma_i (1-p_i)$ for the anti-parallel (AP) configuration. Moreover, a constant spin-flip scattering 
rate, $\gamma_{sf}$, is introduced to model spin relaxation. We also assume infinite on-site Coulomb repulsion to guarantee no double occupation on 
each dot. Therefore, there is a total of $9$ states in this system: no electron in the two QDs, $|00\rangle$, the top QD occupied by one electron 
with spin $\sigma$, $|\sigma 0\rangle$, the bottom QD occupied by one electron with spin $\sigma$, $|0 \sigma\rangle$, and either QD occupied by one 
electron with spin $\sigma$ ($\sigma'$), $|\sigma \sigma'\rangle$. In the sequential-tunneling limit, electronic transport through such double QDs 
can be described by the rate equation:\cite{Cottet,McClure,Dong}    
\bq
\frac{d}{dt} \rho_{\alpha\beta} =\sum_{\alpha'\beta'} M_{\alpha\beta,\alpha'\beta'} \rho_{\alpha'\beta'},\,\,\, 
(\alpha,\beta,\alpha',\beta'={0,\uparrow,\downarrow}), \label{rq}
\eq
in which $\rho_{\alpha\beta}$ is the occupation probability for the state $|\alpha\beta\rangle$. The term 
$M_{\alpha\beta,\alpha\beta}=\sum_{\alpha'\beta'} M_{\alpha'\beta',\alpha\beta}$ gives the total loss rate for the state $|\alpha\beta\rangle$, while 
the term $M_{\alpha\beta,\alpha'\beta'}$ ($\alpha\beta\neq \alpha'\beta'$) gives the total rate for transitions between the two states 
$|\alpha\beta\rangle$ and $|\alpha'\beta'\rangle$. All these terms depend on the spin-flip transition rate and the spin-dependent rates for 
tunneling-in and tunneling-out between a QD $i$ and its lead $\eta i$ ($\eta=L,R$) when the other QD is either empty or occupied, $\Gamma_{\eta i 
\sigma}^{\pm}= \Gamma_{\eta i \sigma} f_{\eta i}^{\pm}(\epsilon_i)$ and $\tilde\Gamma_{\eta i \sigma}^{\pm}= \Gamma_{\eta i \sigma} f_{\eta 
i}^{\pm}(\varepsilon_i+U)$, where $f_{\eta i}^+(\epsilon)=\{1+\exp[(\epsilon - \mu_{\eta i})/k_B T]\}^{-1}$ is the Fermi function of lead $\eta i$ 
with chemical potential $\mu_{\eta i}$ and temperature $T$, and $f_{\eta i}^-(\epsilon_i)= 1- f_{\eta i}^+(\epsilon_i)$. For example, the total loss 
rates for the states $|\sigma 0\rangle$ and $|\sigma \sigma'\rangle$ are $M_{\sigma 0, \sigma 0}= -\sum_{\eta} ( \Gamma_{\eta t \sigma}^{-}+ \tilde 
\Gamma_{\eta b \uparrow}^+ + \tilde \Gamma_{\eta b \downarrow}^+) - \gamma_{sf}$ and $M_{\sigma\sigma', \sigma\sigma'} = -\sum_{\eta} 
(\tilde\Gamma_{\eta t \sigma}^- + \tilde\Gamma_{\eta b \sigma'}^-) - 2\gamma_{sf}$, respectively. Here, a symmetric bias voltage $V_{i}$ is assumed 
to be applied between the left and right leads: $\mu_{L i}=-\mu_{R i}=e V_i/2$.

The steady-state value of $\rho_{\alpha\beta}^0$ under finite bias voltage $V_i$ is obtained by solving $d\rho_{\alpha\beta}/dt=0$ in Eq.~(\ref{rq}). 
Then the spin-dependent currents flowing through the right leads can be evaluated as (we use $\hbar=e=k_B=1$)
\bn
I_{t \uparrow}(t) &=& \Gamma_{Rt \uparrow}^- \rho_{\uparrow 0} -\Gamma_{R t \uparrow}^+ \rho_{00} + \tilde\Gamma_{R t \uparrow}^- 
(\rho_{\uparrow\uparrow} + \rho_{\uparrow \downarrow}) \cr
&& - \tilde\Gamma_{R t \uparrow} (\rho_{0\uparrow} + \rho_{0 \downarrow}), \label{It} \\
I_{b \uparrow}(t) &=& \Gamma_{Rb \uparrow}^- \rho_{\downarrow 0} -\Gamma_{R b \uparrow}^+ \rho_{00} + \tilde\Gamma_{R b \uparrow}^- 
(\rho_{\uparrow\uparrow} + \rho_{\downarrow\uparrow}) \cr
&& - \tilde\Gamma_{R b \uparrow} (\rho_{\uparrow 0} + \rho_{\downarrow 0}), \label{Ib}
\en
and $I_{t \downarrow}$ and $I_{b \downarrow}$ are obtained by interchanging $\uparrow\leftrightarrow \downarrow$ in Eqs.~(\ref{It}) and (\ref{Ib}), 
respectively.
Using the techniques developed in Refs.~\onlinecite{Cottet,McClure,noise,Djuric}, we introduce spin-resolved current matrices for the two circuits 
according to Eqs.~(\ref{It}) and (\ref{Ib}), and then apply them to the steady-state solutions of Eq.~(\ref{rq}) to calculate the average currents 
$I_{t(b)}=\sum_{\sigma}I_{t(b) \sigma}$ and the spin-dependent noise CC $S_{t\sigma,b\sigma'}(\omega)$ defined as
\bq
S_{t \sigma,b \sigma'}(\omega)=2\int_{-\infty}^{+\infty} d\tau e^{i \omega t} \langle \Delta I_{t \sigma}(\tau) \Delta I_{b \sigma'}(0) \rangle ,
\eq
with $\Delta I_{i \sigma}(t)= I_{i\sigma} (t)- \langle I_{i \sigma} \rangle$. The total zero-frequency current CC is $S_{tb}=\sum_{\sigma\sigma'} 
S_{t\sigma, b\sigma'}(0)$.

\begin{figure}[htb]
\includegraphics[height=9cm,width=11cm]{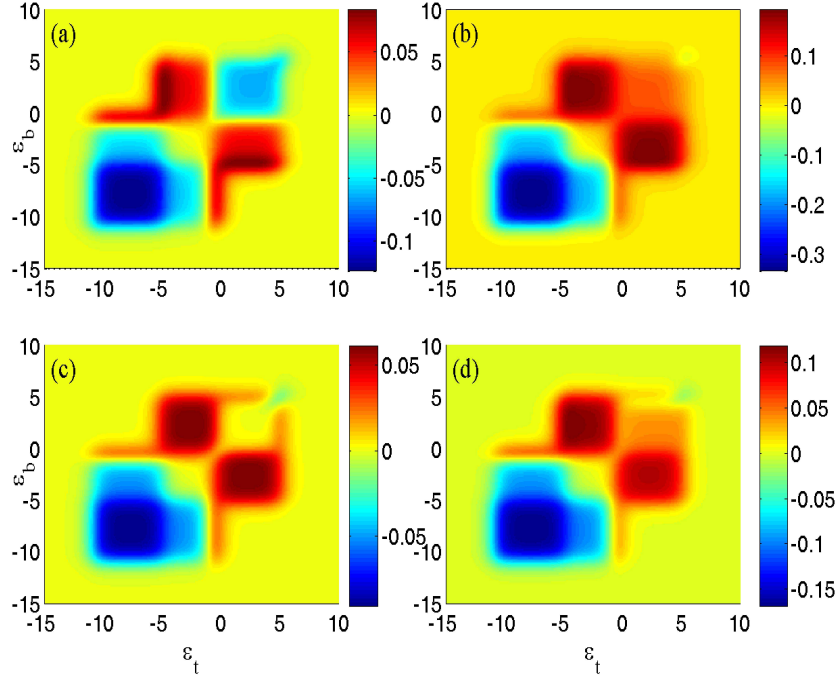}
\caption{(Color online) Zero-frequency cross-correlation dependence on both gate voltages for transport bias voltage $V=10 \gamma$. (a) is plotted 
for paramagnetic leads; (b-c) are for ferromagnetic leads with $p=0.6$ and without spin-flip scattering $\gamma_{sf}=0$. Both circuits are in the P 
configuration (b), the AP configuration (c), and one circuit is in the P configuration and the other one is in the AP configuration (d).}
\label{fig2}
\end{figure}

In the following calculation, we focus on the situation of two identical circuits, $p_t=p_b=p$ and $\gamma_t=\gamma_b=\gamma$. The other parameters 
are: $U=6\gamma$, $T=\gamma/2$ and a given bias voltage $V_t=V_b=V=10\gamma$ (hereafter, in the text and in the figures we use $\gamma$ as the energy 
unit). Figure 2 exhibits the calculated zero-frequency current CC dependence on the energy levels of the two QDs, $\varepsilon_t$ and 
$\varepsilon_b$, for paramagnetic leads (a), and for ferromagnetic leads with strong polarization, $p=0.6$, and no spin-flip scattering 
$\gamma_{sf}=0$ (b-c). In the paramagnetic case, the current CC is, as expected, negative for the occupation regimes, (0,0) and (1,1), while it is 
positive for the regimes, (1,0) and (0,1). It is already known that this positive CC stems from interaction-induced enhancement of the tunneling 
rate.\cite{McClure} To reveal this with greater clarity, we have derived an analytic expression for the current CC for the regime (1,0) (for example, 
setting $\varepsilon_t=-5.5$ and $\varepsilon_b=0$) by assuming an enhanced tunneling-in rate for the bottom QD under the condition that the top QD 
is occupied by an electron, $\tilde \Gamma_{Lb\sigma}^+=x$ ($0<x\leq 1/2$) [correspondingly, the tunneling-out rate is $\tilde 
\Gamma_{Lb\sigma}^-=1-x$; the other rates are either $1$ or $0$]. In the limit of zero temperature, the analytic result is given by
\bq
S_{tb}=-{\frac {2}{27}}\,\gamma\,x\,{\frac { 89\,{x}^{2}+2813\,x-1890 }{ \left( 5\,x+6 \right) ^{3}}}>0 .
\eq   
Surprisingly, we find that the sign of current CC for the transport regime (0,0) is changed in the presence of ferromagnetic electrodes 
[Fig.~2(b-c)], and its largest values are obtained in the case when both two circuits are in the P configuration (P-P configuration). Furthermore, 
the positive CC noise for the transport regimes, (1,0) and (0,1), is enhanced in comparison with those in the case of paramagnetic leads, if either 
of the two circuits is in the P configuration. 
This sign-reversal and the enhancement can be explained in terms of the associated effect of dynamical channel blockade and a mechanism of dynamical 
spin blockade, as suggested by Cottet {\it et al.}.\cite{Cottet}  

Let us focus on the P-P configuration. Here, we assume that up spins are in the majority. Correspondingly, the dwell time of down spins on the dot is 
longer than that of up spins since the down spin tunneling rates (slow electrons) are much lower than those of up spins (fast electrons). Considering 
further the fact that the total number of electrons with up spin, $n_{i\uparrow}$, is equal to that of electrons with down spin, $n_{i\downarrow}$, 
in this configuration, we can infer that every tunneling event of a down-spin electron flowing through a QD must follow several consecutive tunneling 
events of up-spin electrons flowing through the QD, i.e. up spins bunching on both circuits. This indicates that for most time intervals, the sign of 
$\Delta I_{t\uparrow}(\tau)$ is the same as that of $\Delta I_{b \uparrow}(0)$, but is opposite to that of $\Delta I_{b \downarrow}(0)$, which is 
responsible for the occurrence of positive CC between up spins, $S_{t\uparrow,b\uparrow}$(0); but negative CC involves down spins, 
$S_{t\uparrow,b\downarrow}(0)$, for the transport regimes, (0,0), (1,0), and (0,1). We have obtained analytical expressions for the various 
spin-resolved CCs as:
\bn
S_{t\uparrow,b\uparrow}(0) &=& -\frac{2}{125} \gamma (1+p)^2 \frac{\gamma (1-5p) +2 \gamma_{sf}}{2\gamma_{sf} + \gamma (1-p^2)}, \\
S_{t\downarrow,b\downarrow}(0) &=& -\frac{2}{125} \gamma (1-p)^2 \frac{\gamma (1+5p) +2 \gamma_{sf}}{2\gamma_{sf} + \gamma (1-p^2)}, \\
S_{t\uparrow,b\downarrow}(0) &=& S_{t\downarrow,b\uparrow}(0) = \frac{-2\gamma}{125} (1-p^2) \frac{\gamma +2 \gamma_{sf}}{2\gamma_{sf} + \gamma 
(1-p^2)}, \cr
&& 
\en
and for the total CC as:
\bq
S_{tb}(0) = -\frac{8}{125} \gamma \frac{\gamma (1-5p^2) +2 \gamma_{sf}}{2\gamma_{sf} + \gamma (1-p^2)}, \label{snc00}
\eq
in the transport regime (0,0) at zero temperature. Clearly, the positive sign of $S_{tb}(0)$ stems from the up-up CC $S_{t\uparrow,b\uparrow}(0)>0$ 
for sufficiently strong polarization $p>1/\sqrt{5}\approx 0.45$ in absence of spin-flip scattering $\gamma_{sf}=0$ (note that the other spin-resolved 
CCs are all negative). Moreover, retaining spin coherence during tunneling is a necessary condition for such bunching of up spins. We exhibit the 
effect of spin-flip scattering on the zero-frequency current CCs in Fig.~3. It should be noted that spin-flip scattering strongly influences the 
up-up CC once $\gamma_{sf}$ is comparable to the tunneling rate $\gamma$; and in the limit of infinitely high $\gamma_{sf}$, the CCs tend to the 
corresponding values of the paramagnetic case for any value of the polarization and any configuration [see Eq.~(\ref{snc00})]. In particular, 
$S_{tb}(0)$ becomes negative around $\gamma_{sf}\simeq 2\gamma$ in Fig.~3(a). Moreover, a weak polarization, $p$, can not cause sufficiently strong 
up-up spin bunching to overcome the negative CCs, $S_{t\sigma,b\downarrow}(0)$ and $S_{t\downarrow,b\sigma}(0)$, leading to a small negative value of  
$S_{tb}(0)$, as shown in Fig.~3(d) for the results with $p=0.3$. 

\begin{figure}[htb]
\includegraphics[height=9cm,width=11cm]{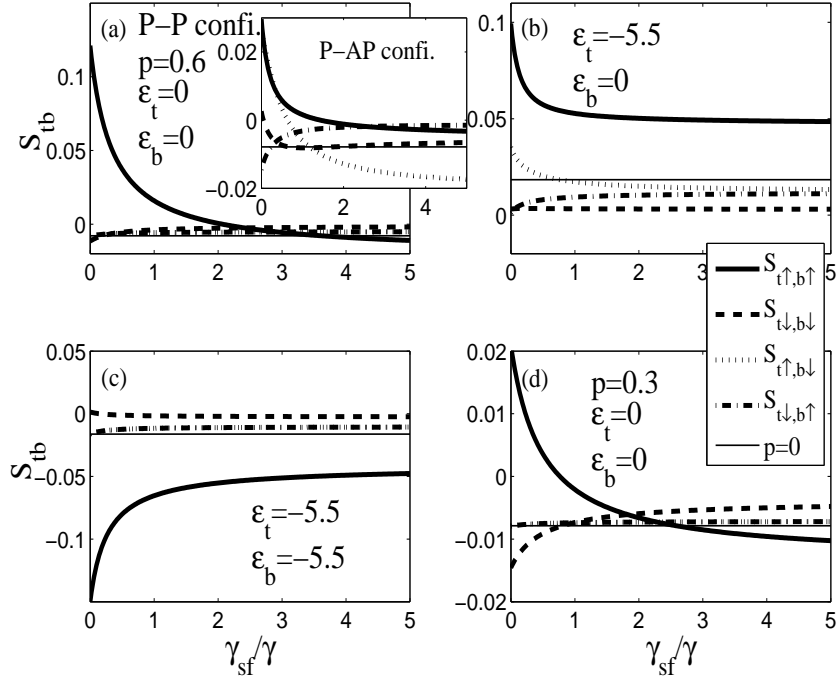}
\caption{Spin-resolved noise cross-correlations as a function of spin-flip scattering, $\gamma_{sf}$, for the case of ferromagnetic leads in the P-P 
configuration. (a-c) are plotted for $p=0.6$ (solid lines), $p=0$ (thin line), and the chosen values of energy levels of the two QDs corresponding to 
the various transport regimes, (0,0), (1,0), and (1,1), respectively; (d) is plotted for $p=0.3$ in the transport regime (0,0). The Inset in (a) 
shows the corresponding results for the transport regime (0,0) with $p=0.6$ in the P-AP configuration.}
\label{fig3}
\end{figure}

Figure 3(b) exhibits the results for the transport regime (1,0). It is obvious that the spin-resolved CCs are all positive due to dynamical channel 
anti-blockade; moreover, the dynamical spin blockade strongly enhances $S_{t\uparrow,b\uparrow}(0)$ and weakly suppresses 
$S_{t\downarrow,b\downarrow}(0)$ and $S_{t\downarrow,b\uparrow}(0)$ in comparison with the results of the paramagnetic case. Therefore, the two 
effects are positively additive and they result in an enhanced positive CC for the transport regimes, (1,0) and (0,1).
For the double occupation case (1,1), inter-dot Coulomb blockade is the dominant mechanism governing transport (over the dynamical spin blockade), 
giving rise to a negative value, but enhanced in magnitude, for the up-up CC [Fig.~3(c)].

On the contrary, the situation is quite different when one of the circuits is in the AP configuration. For the AP case, the fast electron has spin up 
at the input terminal but has spin down at the output terminal, leading to an accumulation of up spins and a rather weak bunching of tunneling events 
associated with down spin at the output terminal.
When the top circuit is in the P configuration while the bottom circuit is in the AP configuration (P-AP configuration), we find very small positive 
values for $S_{t\uparrow,b\uparrow}(0)$ and $S_{t\uparrow,b\downarrow}(0)$, and thus a vanishing $S_{tb}(0)$ for the transport regime (0,0) [inset of 
Fig.~3(a) and Fig.~2(d)]. A similar result is obtained when both circuits are in the AP configuration [Fig.~2(c)]. 

Our results show that a parallel-coupled QD in the P-P configuration is optimal for the occurrence of positive current CC: in the Coulomb 
anti-blockade regimes, (1,0) and (0,1), the maximum value of the CC can be nearly two times higher than that of the paramagnetic case. For instance, 
$S_{tb}(0)\simeq 0.06\gamma$ with $p=0$, while $S_{tb}(0)\simeq 0.18 \gamma$ with $p=0.6$ and $\simeq 0.1 \gamma$ with $p=0.4$. In the Coulomb 
blockade regime, (0,0), the current CC can be $S_{tb}(0)\simeq 0.08 \gamma$ for $p=0.6$ and $\simeq 0.02 \gamma$ for $p=0.5$. 
The sign-change of the current CC between different output terminals due to dynamical spin blockade has not yet been observed experimentally for the 
original setup, a three-terminal QD with ferromagnetic leads.\cite{Cottet} The main difficulty may be due to attaching the third ferromagnetic lead 
to the QD such that its tunnel coupling strength is comparable to that of the others. Our present proposal involves only a simpler element, an 
interacting QD connected to two ferromagnetic leads, which has already been realized experimentally,\cite{fif} and it may provide a more accessible 
setup for the observation of positive CCs originating from dynamical spin blockade.  
      
In conclusion, we have analyzed the gate-voltage dependent CC noise between two output terminals in tunneling through two capacitively coupled QDs 
connected to four independent ferromagnetic electrodes. We find a sign reversal of the zero-frequency CC noise in the dynamical channel blockade 
regime, if the polarization of the electrodes is sufficiently strong. Moreover, in the dynamical channel anti-blockade regimes, positive CC noises 
are obviously enhanced in comparison with the results of paramagnetic leads. These results may be ascribed to the joint effect of dynamical Coulomb 
blockade and dynamical spin blockade, and are found to be quite robust against the spin-flip relaxation.   

This work was supported by Projects of the National Science Foundation of China, Specialized Research Fund for the Doctoral Program of Higher 
Education (SRFDP) of China, and the Program for New Century Excellent Talents in University (NCET). NJMH gratefully acknowledges support by DARPA 
grant No.HR0011-09-1-0008.

\end{document}